**Lethality and centrality in protein networks**

Cell biology traditionally identifies proteins based on their individual actions as catalysts, signaling molecules, or building blocks of cells and microorganisms. Currently, we witness the emergence of a post-genomic view that expands the protein's role, regarding it as an element in a network of protein-protein interactions as well, with a 'contextual' or 'cellular' function within functional modules [1, 2]. Here we provide quantitative support for this paradigm shift by demonstrating that the phenotypic consequence of a single gene deletion in the yeast, *S. cerevisiae*, is affected, to a high degree, by the topologic position of its protein product in the complex, hierarchical web of molecular interactions.

The *S. cerevisiae* protein-protein interaction network we investigate has 1870 proteins as nodes, connected by 2240 identified direct physical interactions, and is derived from combined, non-overlapping data [3, 4] obtained mostly by systematic two-hybrid analyses [3]. Due to its size, a complete map of the network (Fig. 1a), while informative, in itself offers little insight into its large-scale characteristics. Thus, our first goal was to identify the architecture of this network, determining if it is best described by an inherently uniform exponential topology with proteins on average possessing the same number of links, or by a highly heterogeneous scale-free topology with proteins having widely different connectivities [5]. As we show in Fig. 1b, the probability that a given yeast protein interacts with *k* other yeast proteins follows a power-law [5] with an exponential cutoff [6] at $k_c \cong 20$, a topology that is also shared by the protein-protein interaction network of the bacterium, *H. pylori* [7]. This indicates that the network of protein interactions in two separate organisms forms a highly inhomogeneous scale-free network in which a few highly connected proteins play a central role in mediating interactions among numerous, less connected proteins.

An important known consequence of the inhomogeneous structure is the network's simultaneous tolerance against random errors coupled with fragility against the removal of the most connected nodes [8]. Indeed, we find that random mutations in the genome of *S. cerevisiae*, -modeled by the removal of randomly selected yeast proteins-, do not affect the overall topology of the network. In contrast, when the most connected proteins are computationally eliminated, the network diameter increases rapidly. This simulated tolerance against random mutation is in agreement with systematic mutagenesis studies, which identified a striking capacity of yeast to tolerate the deletion of a substantial number of individual proteins from its proteome [9,10]. Yet, if indeed this is due to a topological component to error tolerance, on average less connected proteins should prove less essential than highly connected ones. To assess this hypothesis, we rank ordered all interacting proteins based on the number of links they have and correlated this with the phenotypic effect of their individual removal from the yeast proteome. As shown in Fig. 1c, the likelihood that removal of a protein will prove lethal clearly correlates with the number of interactions the protein has. For example, while proteins with five or less links constitute ~93% of the total number of proteins we find that only ~21% of them are essential. In contrast, only ~0.7% of the yeast proteins with known phenotypic profile have more than 15 links but single deletion of ~62% of these proves lethal. This implies that highly-connected proteins with a central role in the network's architecture are three times more likely to prove essential than proteins with low number of links to other proteins.

The simultaneous emergence of an inhomogeneous structure in both metabolic-[5,11] and protein interaction networks indicates the evolutionary selection of a common large scale structure of biological networks, and strongly suggests that future systematic protein-protein interaction studies in other organisms will uncover an essentially identical protein network topology. The correlation between the connectivity and indispensability of a given protein confirms that despite the importance of individual biochemical function and genetic redundancy, the robustness against mutations in yeast is also derived from the organization of interactions and topologic position of individual proteins [12]. Thus, a better understanding of cell dynamics and robustness will be obtained from integrated approaches that simultaneously incorporate the individual and contextual properties of all constituents of complex cellular networks.

H. Jeong[1], S. P. Mason[2], A.-L. Barabási[1] and Z. N. Oltvai[2]

[1] Department of Physics, University of Notre Dame, Notre Dame, IN 46556 and

[2] Department of Pathology, Northwestern University Medical School, Chicago, IL 60611

E-mail: alb@nd.edu, zno008@nwu.edu

**Figure Legend**   Characteristics of the yeast proteome.

(**a**) Map of protein-protein interactions. The largest cluster, which contains ~78% of all proteins is shown. The color of a node signifies the phenotypic effect of removing the corresponding protein (red=lethal, green=non-lethal, orange=slow growth, yellow=unknown) (**b**) Connectivity distribution *P(k)* of interacting yeast proteins giving the probability that a given protein interacts with *k* other proteins. The exponential cutoff [6] indicates that the number of proteins with more than 20 interactions is slightly less than expected for pure scale-free networks. In the absence of data on the link directions, all interactions have been considered bi-directional. The parameter controlling the short length scale correction has value $k_0 \approx 1$.   (**c**) The fraction of essential proteins with exactly *k* links versus their connectivity *k* in the yeast proteome. The list of 1572 mutants with known phenotypic profile was obtained from the Proteome database [13]. Detailed statistical analysis, including *r=0.75* value for Pearson's linear correlation coefficient, demonstrates a positive correlation between lethality and connectivity. For additional details see http://www.nd.edu/~networks/cell.

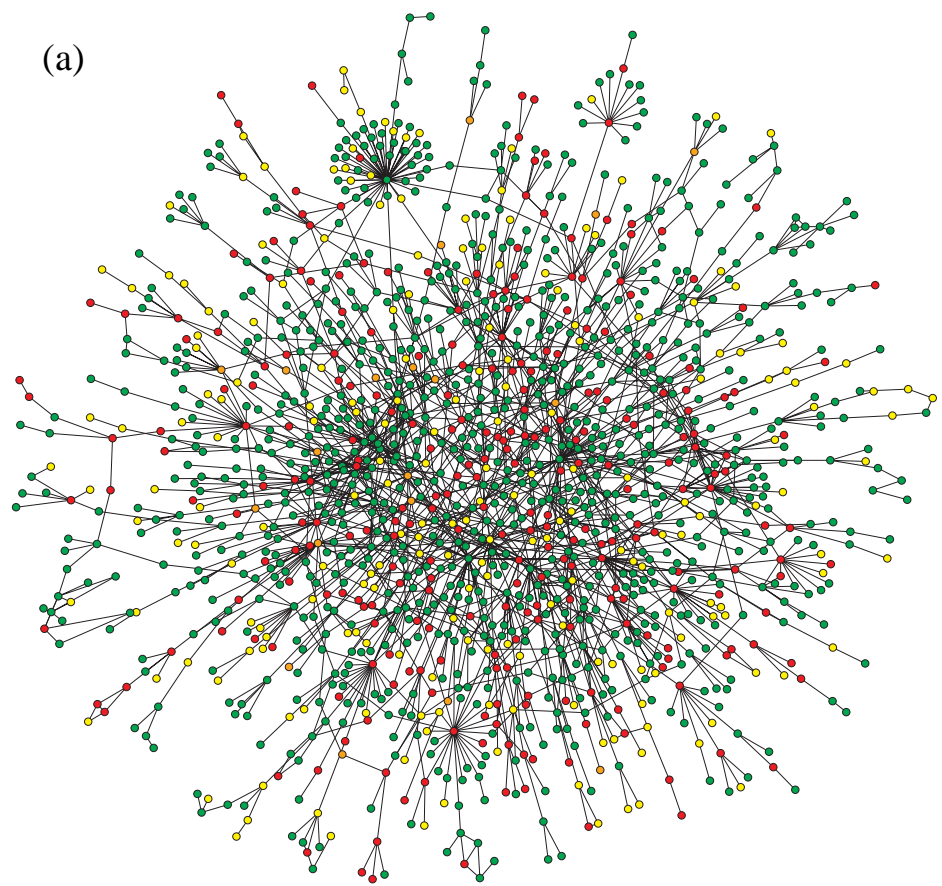
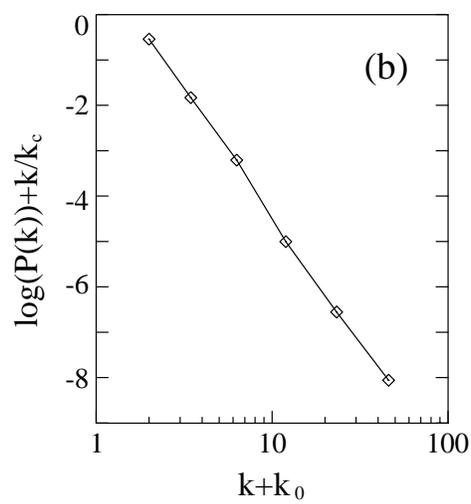
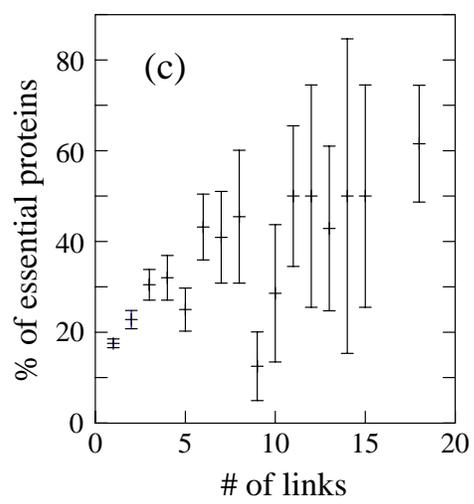